\documentclass[longauth]{aa}
\usepackage{amssymb,amsmath,amsfonts}
\usepackage{graphicx}
\usepackage{color}

\usepackage{xcolor}

\newcommand{\me}{\, {\rm M}_{\oplus}}

\newcommand{\orcid}[1]{\protect\href{https://orcid.org/#1}{\protect\includegraphics[width=8pt]{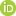}}}

\newcommand{\er}{\xi_r}
\newcommand{\es}{\xi_s}

\definecolor{cadmiumgreen}{rgb}{0.0, 0.42, 0.24}

\newcommand{\be}{\begin{equation}}
\newcommand{\ee}{\end{equation}}

\usepackage{amsmath}
\usepackage{bbold}
\usepackage{tikz}
\usepackage{bm}

\usepackage[pdftex,colorlinks=true,linkcolor=blue,citecolor=blue,urlcolor=blue]{hyperref}

\begin{document}
\nolinenumbers
\title{Insights into planetary formation pathways from the \texttt{resonantstate} databases
%
}
\subtitle{
II. Testing models against observed architectures
}

\author{
A. Leleu\inst{1}\orcid{0000-0003-2051-7974},
M. Goldberg\inst{2}\orcid{0000-0003-3868-3663},
S. Marques\inst{3,4}\orcid{0009-0007-3825-7791},
S. Ataiee\inst{5,6}\orcid{0000-0003-2594-3454},
S. Shibata\inst{7}\orcid{0000-0002-5418-6336},
M. Rosenqvist\inst{1}\orcid{0009-0009-6885-7450},
J.-B. Delisle\inst{1}\orcid{0000-0001-5844-9888},
Y. Ishida\inst{8}\orcid{0009-0004-2879-1462},
R. Mardling\inst{1,9}\orcid{0000-0001-7362-3311},
J. Couturier\inst{1,10}\orcid{0000-0001-8897-9978},
Y. Eyholzer\inst{1}\orcid{0009-0004-3041-2249},
D. C. Fabrycky\inst{11}\orcid{0000-0003-3750-0183},
A. C. Petit\inst{2}\orcid{0000-0003-1970-1790},
E. Agol\inst{12}\orcid{0000-0002-0802-9145},
S. Hadden\inst{13}\orcid{0000-0002-1032-0783},
A. Izidoro\inst{14}\orcid{0000-0003-1878-0634},
G. Pichierri\inst{15}\orcid{0000-0003-3622-8712},
Y. Alibert\inst{3,4},
E. Kokubo\inst{16}\orcid{0000-0002-5486-7828},
C. Mordasini\inst{3,4}\orcid{0009-0007-3825-7791},
J.M. Almenara\inst{1}\orcid{0000-0003-3208-9815},
L. Borsato\inst{17}\orcid{0000-0003-0066-9268},
R. Burn\inst{2}\orcid{0000-0002-9020-7309},
F. Dai\inst{18}\orcid{0000-0002-8958-0683},
A. Emsenhuber\inst{19}\orcid{0000-0002-8811-1914},
I.-G. Jiang\inst{20},
J. Korth\inst{1,21,22},
A.-S. Libert\inst{23}\orcid{0000-0003-2682-2793},
M. MacDonald\inst{24}\orcid{0000-0003-2372-1364},
K. Masuda\inst{25}\orcid{0000-0003-1298-9699},
S. Millholland\inst{26,27},
D. Ragozzine\inst{28},
S. Udry\inst{1},
M.-T. Wang\inst{29,30}\orcid{0000-0003-3015-6455},
T. G. Wilson\inst{31}\orcid{0000-0001-8749-1962},
D. A. Yahalomi\inst{27}\orcid{0000-0003-4755-584X},
A. Alnajjarine\inst{32}\orcid{0009-0005-5366-3371},
A. C. M. Correia\inst{32,33}\orcid{0000-0002-8946-8579},
R. Eltschinger\inst{3,4}\orcid{0009-0001-7351-3232},
Z.-C. Hu\inst{34}\orcid{0009-0000-6461-5256},
J. Laskar\inst{32},
D. Ségransan\inst{1},
D. Tamayo\inst{35}\orcid{0000-0002-9908-8705},
L.-C. Yeh\inst{36}
}

\institute{
 Observatoire astronomique de l'Université de Genève, Chemin Pegasi 51, 1290 Versoix, Switzerland 
\and Université Côte d’Azur, Observatoire de la Côte d’Azur, CNRS, Laboratoire Lagrange, UMR7293, Boulevard de l’Observatoire, 06304, Nice Cedex 4, France 
\and Space Research and Planetary Sciences, Physics Institute, University of Bern, Gesellschaftsstrasse 6, 3012 Bern, Switzerland  
\and Center for Space and Habitability, University of Bern, Gesellschaftsstrasse 6, 3012 Bern, Switzerland 
\and Department of Physics, Faculty of Sciences, Ferdowsi University of Mashhad, Mashhad, 91775-1436, Iran 
\and Research Center for Modeling and Computational Sciences, Ferdowsi University of Mashhad, Mashhad, Iran 
\and Department of Earth, Environmental and Planetary Sciences, MS 126, Rice University, Houston, TX 77005, USA 
\and Department of Astronomy, The University of Tokyo, 7-3-1 Hongo, Bunkyo-ku, Tokyo 113-0033, Japan 
\and School of Physics \& Astronomy, Monash University, Clayton 3800, Australia 
\and LTE, Observatoire de Paris, Univ. PSL, Sorbonne Univ., Univ. de Lille, LNE, CNRS, 61 Av. de l’Observatoire, 75014 Paris, France  
\and Department of Astronomy \& Astrophysics, University of Chicago, 5640 S. Ellis Ave., Chicago, IL 60637, USA  
\and Astronomy Department and Virtual Planetary Laboratory, University of Washington, Seattle, WA 98195, USA  
\and Canadian Institute for Theoretical Astrophysics, 60 St George St Toronto, ON M5S 3H8, Canada  
\and Department of Earth, Environmental and Planetary Sciences, 6100 MS 126, Rice University, Houston, TX 77005, USA  
\and Dipartimento di Fisica, Università degli Studi di Milano, Via Celoria 16, 20133 Milano, Italy  
\and Center for Computational Astrophysics, National Astronomical Observatory of Japan, Osawa, Mitaka, Tokyo 181-8588, Japan  
\and INAF, Osservatorio Astronomico di Padova, Vicolo dell’Osservatorio 5, 35122 Padova, Italy  
\and Institute for Astronomy, University of Hawai`i, 2680 Woodlawn Dr, Honolulu, HI, 96822, USA  
\and Physics Institute, University of Bern, Sidlerstrasse 5, 3012 Bern, Switzerland  
\and Department of Physics and Institute of Astronomy, National Tsing Hua University, Hsinchu City 30013, Taiwan  
\and Instituto de Astrof\'{i}sica de Canarias (IAC), 38205 La Laguna, Tenerife, Spain  
\and Departamento de Astrof\'isica, Universidad de La Laguna (ULL), E-38206 La Laguna, Tenerife, Spain  
\and naXys, Departement of Mathematics, Rue de Bruxelles 61, 5000 Namur, Belgium  
\and Department of Physics, The College of New Jersey, 2000 Pennington Rd, Ewing, NJ 08628, USA  
\and Department of Earth and Space Science, Graduate School of Science, Osaka University, Toyonaka, Osaka 560-0043, Japan  
\and Department of Physics, Massachusetts Institute of Technology, Cambridge, MA 02139, USA  
\and Kavli Institute for Astrophysics and Space Research, Massachusetts Institute of Technology, Cambridge, MA 02139, USA  
\and Brigham Young University Department of Physics \& Astronomy, N283 ESC, Brigham Young University, Provo, UT 84602, USA  
\and School of Astronomy and Space Science, Nanjing University, Nanjing 210023, China.  
\and Key Laboratory of Modern Astronomy and Astrophysics, Ministry of Education, Nanjing, 210023, People’s Republic of China  
\and Department of Physics, University of Warwick, Gibbet Hill Road, Coventry CV4 7AL, UK  
\and LTE, Observatoire de Paris, Univ. PSL, Sorbonne Univ., CNRS, 77 Av. Denfert-Rochereau, 75014 Paris, France  
\and CFisUC, Departamento de F\'isica, Universidade de Coimbra, 3004-516 Coimbra, Portugal 
\and Department of Astronomy, Tsinghua University, Beijing 10084, China  
\and Department of Physics, Harvey Mudd College, 301 Platt Blvd, Claremont, CA 91711, USA  
\and Institute of Computational and Modeling Science, National Tsing Hua University, Hsinchu City 30013, Taiwan
}

 \abstract
  { 
The architecture of planetary systems near or in mean motion resonances encodes information on their formation pathways. Two observations notably challenge formation models: resonant systems are rare, and near-resonant pairs pile up just wide of exact commensurability.
To understand which physical processes took place during the formation and evolution of the observed population, we developed two \texttt{resonantstate} databases: an observational database—presented in Paper I—and a simulation database. The latter compiles the evolution of over ten thousand simulated systems which are evolved via various physical processes, including accretion, planet-disk interactions, N-body evolution, and tidal dissipation. We test the physical processes against the observed population by projecting both databases onto the second fundamental model of resonance (Paper I).

For systems far enough from the star to not be strongly affected by planetary tides, we find that the small fraction of observed pairs that are in resonance has a resonant state qualitatively reproduced by hydrodynamical simulations and a subset of the population synthesis models, which supports the disk parameters and models that were used. 
For pairs outside of resonance, simulations dominated by dissipative processes do not reproduce the observed excited eccentricities, while simulations dominated by scattering and collisions do not reproduce the over-abundance of pairs wide of the commensurability. 
Instead, models that undergo a phase of resonant capture, followed by post-disk instabilities, qualitatively reproduce both of these observables.
The closest-in systems, which we showed in Paper I to be farther from commensurability and less eccentric, can be explained by dynamical evolution similar to the farther-out systems followed by planetary tides.
Finally, population synthesis models—previously thought to broadly over-produce near-resonant systems—match the observed near-resonant fraction once their resonance-breaking rate is extrapolated to 5~Gyr; however, pairs inside a resonance are still over-abundant by a factor $\sim\!3$. In order to corroborate these results, and broaden the scope, we welcome new simulations of the formation and evolution of multi-planetary systems to be added to the \texttt{resonantstate} database.
     }


\titlerunning{Planet formation from resonant states}

\authorrunning{Leleu et al.}

\maketitle
\nolinenumbers

\section{Introduction}

The observed architectures of planetary systems are the byproducts of numerous physical processes which occur during their formation in the proto-planetary disk, as well as millions to billions of years of subsequent evolution after disk dispersal.
Among the possible architectures, systems hosting pairs of planets whose period ratio is close to a low-order commensurability are of particular interest. 
Such configurations form naturally because convergent orbital migration and eccentricity damping, caused by planet-disk interactions, lock planets into mean-motion resonant equilibria. \citep{GoTre1979,Weidenschilling1985,NePa2003,Masset2006,BatPet2023}. 

The discovery of hundreds of transiting multi-planetary systems by the Kepler mission has provided a statistical view of these systems and enabled comparison with theoretical models of their origin.
Early work identified trends in the distribution of period ratios of near-resonant systems, specifically an excess of pairs wide of exact commensurability and a dearth interior to it \citep{Lee2013,Fabrycky2014,Delisle2014letter}.
More recent studies have identified correlations of proximity to resonance with other quantities, such as the planetary densities and mutual inclinations \citep{Millholland2020,Millholland2022,Leleu2024b}. Several studies have also found evidence that young stars, as identified by cluster membership, gyrochronology, and lithium abundances, have a higher occurrence of transiting near-resonance planet pairs \citep{Dai2024} and transit timing variations, likely caused by near-resonant companions \citep{LopezMurillo2026}. \cite{Hamer2024} found that hosts of near-resonant Kepler planets have a smaller galactic velocity dispersion than the overall sample, indicating their youth, but these particular results have been disputed by \cite{Zhu2026}.

From a theoretical perspective, the existence of resonant systems typically implies that a dissipative process, almost certainly orbital migration, played a major role in their formation and assembly \citep{LePe2002}. However, almost all theoretical models incorporating migration suffer from two major discrepancies when compared to data. The first problem is the relative rarity of observed resonant systems in Kepler when compared to planet formation models that account for orbital migration. To reconcile this resonant overproduction problem, \cite{Mills2016,Izidoro2017} proposed the `breaking-the-chains' mechanism. In this model, the majority of systems indeed form resonant chains in the disk, but most of them become dynamically unstable during the Gyrs of dynamical evolution following the dispersal of the gaseous disk \citep{TePa2007,OgIda2009,Cossou2014,Izidoro2017}. As such, the dynamical history of exoplanet systems may resemble that of the giant planets in the Solar System \citep{Mo2007, Thommes2008}. The breaking-the-chains model is supported by emerging trends in the bulk planet density and host star age distributions of near-resonant systems \citep{Leleu2024b,Dai2024}, although the precise mechanisms to disrupt the chains are still being explored \citep[e.g.][]{GoPe2026,HaWu2026,Choksi2026,Liu2026}. We also note that some studies have proposed mechanisms to avoid forming resonances during migration, including turbulence (\citealp{Adams2008}, but see \citealp{Batygin2017}), inviscid disks \citep{McNally2019}, and resonant overstability (\citealp{GoSc2014}, but see \citealp{DeBa2015}).

The second problem is that most of the observed near-resonant planet pairs are notably wide of exact commensurability \citep{Lissauer2011,Fabrycky2014}. Although first-order mean motion resonances are asymmetric, so that the equilibrium state lies wide of commensurability \citep{Pichierri2018}, the magnitude of the observed offset is much larger than what is predicted by standard theories of migration. The effect is particularly noticeable in $3+$ planet systems where three-body resonances are preserved \citep{Charalambous2023}. \cite{LithwickWu2012} and \cite{Batygin2013} proposed that tidal dissipation within the planets, operating over the lifetime of the system, was responsible for ``repelling'' the planet pairs. Although there is evidence for tidal evolution in the eccentricity distribution and in the magnitude of the shift from exact commensurability for pairs whose inner period is less than 5 days \citep[][]{Delisle2014letter,HaLi2017,GRSW1}, \cite{GRSW1} showed that the over-abundance of pairs wide of commensurability exists and is similar for pairs at larger separation from their host star that retained non-negligible eccentricities. This excess is therefore unlikely to be mainly due to tides.
In fact, other mechanisms that dissipate orbital energy can have a similar effect by evolving along the resonant equilibrium curve, including scattering of small bodies \citep{ChatterjeeFord15,Wu2024} and very dissipative disks \citep{Ramos2017}.

The principal challenge in testing these hypotheses is that for many observed close-in planetary systems, transits and radial velocities provide 
only little dynamical information. The signatures of evolution in and out of resonance may be preserved in the orbital architectures of the planetary systems and the planetary structures themselves. Detailed characterization of resonant and near-resonant systems is thus a powerful way to understand both their origins as well as the origins of many other planetary systems, which may have undergone a transient resonant phase.
\defcitealias{GRSW1}{RS1}
These architectures can notably be characterized by analyzing transit timing variations (TTVs) induced by near-resonant planetary systems \citep{Agol2005,NeVo2016,AgolFabrycky2025}. In the first paper of the resonant state series \citep[][hereafter \citetalias{GRSW1}]{GRSW1}, we present the \texttt{resonantstate} observational database that contains samples of TTV analyses of over 600 multi-planetary systems. We also show how these samples can be used to constrain the `resonant state' of observed pairs near any first-order mean motion resonance by estimating scale-free quantities that describe their resonant interactions.  

In this second paper, we present the \texttt{resonantstate} simulation database, which gathers numerical simulations of the formation and evolution of multi-planetary systems under various physical processes. We describe this database in \S\ref{sec:databases}.
In \S\ref{sec:framework}, we recall the framework we use to compare the resonant state of pairs of planets. In \S\ref{sec:comparison}, we showcase the use of the databases by comparing the resonant state of observed systems to a diverse set of simulated ones. We discuss which observational trends are best reproduced by which physical processes in \S\ref{sec:discussion}. We then summarize our results and offer future directions in \S\ref{sec:conclusion}.

\section{The simulation database}
\label{sec:databases}

To facilitate the comparison of the architecture of observed and simulated systems, we created two publicly available databases in which the osculating astrocentric orbital elements, planetary masses, and other relevant quantities of both observed and simulated systems are stored in the same format. The databases can be explored using the \texttt{resonantstate} package.\footnote{\url{https://github.com/adleleu/resonantstate}} 
The description of the observational database and the general details of the databases and how to use them are provided in \citetalias{GRSW1}.

The simulation database stores the temporal evolution of the simulations that can be accessed using metadata tables. These are broken down into two layers: a run metadata table,\footnote{\url{https://dace.unige.ch/resonantStateSimulationsCatalog/}} and simulation metadata tables. 
A run is defined as a set of simulations which have identical physical assumptions and models, with different initial conditions (e.g. randomized, grid-based, etc.). 

The version of the database used for this study has 21 runs from 7 teams. These runs include 2D hydrodynamical simulations of pairs or triplets of planets at the inner edge of a proto-planetary disk from \cite{Ataiee2021}; pairs of planets evolving in a laminar or turbulent 1D disk from \cite{GoBa2023}; pairs evolving due to the presence of remaining planetesimal disks \citep{GoBa2023}; systems evolving through tides by J. Couturier, A. Alnajjarine and \cite{GoBa2023}. 

The database also includes a run of \textit{N}-body simulations from \citet{Ishida&Kokubo2026}, which adopt the same setup as \citet{Kokubo+2025}. In this run, the evolution is dominated by planet-planet scattering and collisions.
Initial protoplanet systems consist of $15$ protoplanets of equal mass, each with a mass of $1.82\me$, and a density of $3$~g~cm$^{-3}$, and an orbital separation of $10 r_\mathrm{H}$, located between 0.1\, au and 0.3\, au, where $r_\mathrm{H}$ is the mutual Hill radius of the protoplanets. The distribution of the initial values of the eccentricities is a Rayleigh distribution with a dispersion of 0.01, and the initial inclinations also follow a Rayleigh distribution, with a dispersion of 0.005. 
The systems are integrated with a modified Hermite scheme for planetary \textit{N}-body simulations \citep{Kokubo+1998, Kokubo&Makino2004} with block timesteps \citep{Makino1991} for $5\times10^{8}$ orbital periods of the innermost protoplanet under the assumption of perfect accretion.
For details on the simulation, see \citet{Kokubo+2025}.

Finally, the database includes several runs from two population synthesis models. The first population synthesis is the NGPPS Bern model. These simulations consistently evolve a viscously accreting gas disc and the dynamical state of a planetesimal disc along with a given number of protoplanets that concurrently accrete solids and gas.
An N-body integrator which includes forces due to gas-driven migration and damping was used to track the dynamical interaction of the protoplanets \citep{Emsenhuber2021}.
The initial conditions for the population synthesis were chosen to best represent protoplanetary disc demographics \citep{Emsenhuber2021b}.
N-body interactions are usually tracked for 20 Myr, but have been extended to 100 Myr for the nominal population NG76 \citep{Emsenhuber2023,Emsenhuber2025}.


The second population synthesis model combines several sets of simulations from the breaking-the-chains framework. Both \cite{Izidoro2022} and \cite{Shibata2025} modeled the formation of close-in super-Earths and sub-Neptunes by incorporating disk evolution, planetary growth, migration, resonant-chain formation, and post-gas dynamical instabilities. The main difference between these studies lies in the assumed accretion paradigm. In \cite{Izidoro2022}, planets grow predominantly through pebble accretion, with rocky embryos in the inner disk forming and growing efficiently from a pebble-dominated reservoir. In contrast, \cite{Shibata2025} adopted a two-ring model in which rocky embryos form in an inner planetesimal ring near the silicate sublimation line and grow primarily through planetesimal accretion, while icy embryos beyond the snowline grow predominantly through pebble accretion. Thus, the two-ring model retains the same broad dynamical framework but connects the resulting planetary architectures to distinct rocky and icy growth reservoirs.

The simulation data presented in this work are drawn from three models: Model C from \cite{Izidoro2022} and two models from \cite{Shibata2025} with initial inner planetesimal-ring masses of 3 and 6 $M_\oplus$ and disk lifetimes of 3 and 2 Myr, respectively. We selected these models because they successfully reproduce the observed planetary radius distribution. In addition, as demonstrated in \cite{Shibata2025}, these models reproduce other key observational properties, including the period-ratio distribution, planetary multiplicity, and intra-system size uniformity.

The final state of these simulations can be compared to the observed architecture of planetary systems. In this study, we focus on the architecture of resonant and near-resonant pairs of exoplanets.

\section{Framework for the study of resonant and near-resonant pairs}
\label{sec:framework}
\subsection{The second fundamental model for resonance}
\label{sec:SFMR}
\begin{figure}
\begin{center}
\includegraphics[width=0.99\linewidth]{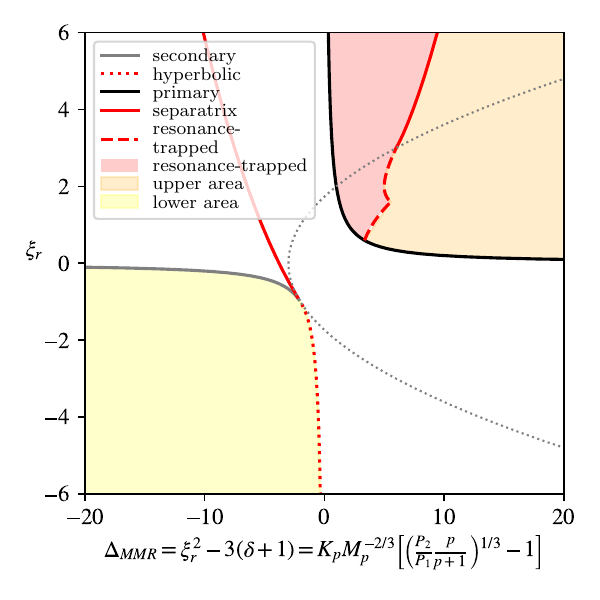}
\caption{Phase space of the SFMR in the ($\Delta_{\rm MMR}$, $\er$) plane.
The black and grey solid curves show the equilibria of the phase space, while the solid red line shows the boundary of the formally resonant domain (red area above $\delta=0$ shown as a dotted grey curve).
`Upper' (resp. `lower') refers to orbits above (resp. below) the exact commensurability. The white areas are parts of the phase space that are redundant with the colored ones and are therefore not used.
%
}
\label{fig:SFM_anarea} 
\end{center}
\end{figure}


In section 3 of \citetalias{GRSW1}, we have highlighted how the second fundamental model of resonance \citep[SFMR,][]{HenLe1983} is a powerful tool to compare the resonant state of any pair of planets near any first-order MMR; and is particularly well suited to systems characterized by TTVs. In this section, we briefly re-introduce the scale-free parameters of the SFMR that we will use to compare the resonant state of observed systems with simulated ones in the rest of the paper, and we encourage the reader to go through section 3 of \citetalias{GRSW1} for more details.

The projection onto the SFMR of osculating orbital elements and masses of a pair of planets near a first order MMR can be done under some conditions: the system is supposed to have small mutual inclinations and eccentricities (first order). The pair also needs to not have a nearby massive perturber. The applicability is described further in section 3.4 of \citetalias{GRSW1}. We will notably discard pairs whose total planetary mass is less than twice the mass of a perturber with a period ratio of less than 3.5 with either planet.

Under these conditions, for a given pair we first define three constant quantities: the resonant parameter $\delta$, which depends on the angular momentum deficit of the pair and describes the phase space of the SFMR; the rescaled resonant eccentricity $\er$, which, combined with $\delta$, defines the position of the pair in that phase space (linked to an energy curve of the SFMR). Finally, $\es$, which is the rescaled secular eccentricity that describes the amplitude of the eccentricities that are not involved in the resonant dynamics. $\er$ and $\es$ scale linearly with the eccentricity of the planets. For a pair of super-Earths near a 3:2 MMR, they are comparable to the planets' eccentricities multiplied by $\sim\!100$. 

Then, following \cite{Lin2025} and \citetalias{GRSW1}, we combine $\delta$ and $\er$ to define the distance to commensurability:
\begin{equation}
\begin{aligned}
\Delta_{\rm MMR} &= \er^2 - 3(\delta+1) = K_p M_p^{-2/3} \left[ \left(\frac{P_2}{P_1} \frac{p}{p+1} \right)^{1/3} -1 
\right], 
\label{eq:DMMR}
\end{aligned}
\end{equation}
The former expression highlights that this constant quantity is scale-free, while the latter shows that it depends on the masses of the planets ($M_p=2m_1/m_\star$ when $m_1=m_2$), and on $P_1$ and $P_2$, which are instantaneous periods reached during the resonant evolution of the pair \citepalias[see][]{GRSW1}. The coefficient $K_p$, which depends on the considered MMR, is given in Table \ref{tab:Kp}.

\begin{table}[!h]
\centering
\caption{$K_p$ used to compute $\Delta_{\rm MMR}$ and $\Delta_{p+1:p}$.}\label{tab:Kp}
\begin{tabular}{c|c c c c c c}
    p & $1$ & $2$   & $3$ & $4$ & $5$ & $6$ \\
    \hline
$K_p$ & $5.64$ & $3.56$ & $3.18$ & $2.92$ & $2.73$ & $2.58$
\end{tabular}
\end{table}

In the ($\Delta_{\rm MMR},\er$) plane we define the resonant area in red, the upper (resp. lower) area $P_2/P_1>(p+1)/p$ (resp. $P_2/P_1<(p+1)/p$). In order to discuss the distance of a pair of planets to the relevant equilibrium for a given value of $\delta$, we write:
\begin{equation}
 \Delta \er = \er-\xi_{ell}(\delta),
\end{equation}
where $\xi_{ell}(\delta)$ is the value of $\er$ at the elliptic point (primary elliptic point for pairs in resonance or $P_2/P_1>(p+1)/p$, and secondary otherwise) following the curve of constant $\delta$ of the pair, and $\Delta \er$ the distance to that equilibrium.
$\Delta \er=1$ when the separatrix appears for $\delta=0$. Finally, following \citetalias{GRSW1} we define the resonance-trapped domain which adds to the formally resonant domain an extension with $\Delta \er=1$ down to $\delta=-2$ (red dashed curves in Fig. \ref{fig:SFM_anarea} a and b).

\subsection{Pair selection}
\label{sec:selection}

In order to attribute a pair to a given MMR, we introduce the distance to commensurability:
\begin{equation}
\begin{aligned}
\Delta_{p+1:p}=M_p^{2/3}\Delta_{\rm MMR} = K_p  \left[ \left(\frac{P_2}{P_1} \frac{p}{p+1} \right)^{1/3} -1 
\right]. 
\label{eq:Dpp1p}
\end{aligned}
\end{equation}
A pair is attributed to the $p\!+\!1\!:\!p$ MMR that minimizes $|\Delta_{p+1:p}|$. Since first order MMRs get closer in period ratios as $p$ gets larger, we assign a pair to resonance $p$ only if $p$ is also the closest resonance for the mirrored distance $-\Delta_{p+1:p}$. This gives a symmetric range around each commensurability \citepalias[see][]{GRSW1}.

\subsubsection{Observed systems}
\label{sec:observed}

For the observed population, the pair selection is identical to \citetalias{GRSW1} : We select pairs of planets that pass the criteria developed in \S 3.3 and 3.4 of that paper: pairs that land in symmetrical domains around $\Delta_{p+1:p}$, no neighbouring massive planets, with an outer orbital period of less than 70 days. 
When several analyses are available in the database, we give priority to robust solutions, then to photo-dynamical analysis over the analysis of pre-extracted transit timings. We obtain 442 observed pairs, of which 56 are considered robust 
\citep{2015MNRAS.453.2644A,2016A&A...595L...5A,MacDonald2016,Mills2016,HaLi2017,2018A&A...615A..90A,2018MNRAS.478..460A,2021A&A...655A..66L,Agol2021,Leleu2022,2022A&A...663A.134A,Leleu2023,2023A&A...675A.115K,2023AJ....165...33D,2023NatAs...7..206P,Leleu2024,2024A&A...689A..52B,2024AJ....168...31W,Masuda2024,2024ApJ...971L..28K,2025PSJ.....6..267L,2025AJ....169..243J,Leleu2026,Ragozzine2026}. See \citetalias{GRSW1} for more details.

\subsubsection{Simulated systems}

In addition to the selection criterion of the observed system, simulations that were set at 1AU for numerical reasons but were meant to reproduce close-in system were considered (e.g. hydrodynamical simulations at the inner disk edge). We discarded planets with masses lower than one Earth mass in population synthesis models, as the simulations contain numerous moon mass embryos that we do not wish to compare to the observed population. Finally, only pairs that would generate peak-to-peak TTVs of more than 30 min over four years are considered for \S \ref{sec:reso} and \S \ref{sec:xlibphase} to discard near-resonant simulated systems that would have been unlikely to be characterized by TTVs.

\section{Comparison of observed and simulated systems}
\label{sec:comparison}

\subsection{Simulated systems}
\label{sec:simu}

\begin{figure}
\begin{center}
\includegraphics[width=0.99\linewidth]{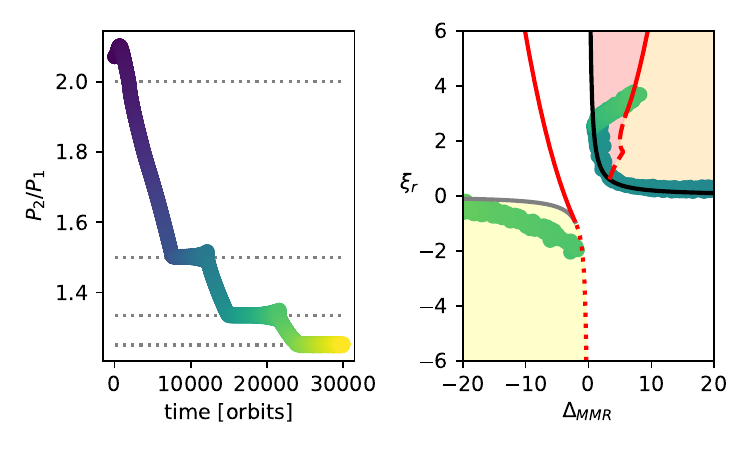}
\caption{Evolution of a pair of planets at the inner edge of the protoplanetary disk through 2D hydrodynamical simulations from \cite{Ataiee2021}. Left panel shows the evolution of the period ratio vs time, while the right panel focuses on the evolution through the 4:3 MMR. The colors indicate the time. 
}
\label{fig:hydro} 
\end{center}
\end{figure}

\begin{figure*}
\begin{center}
\includegraphics[width=0.99\linewidth]{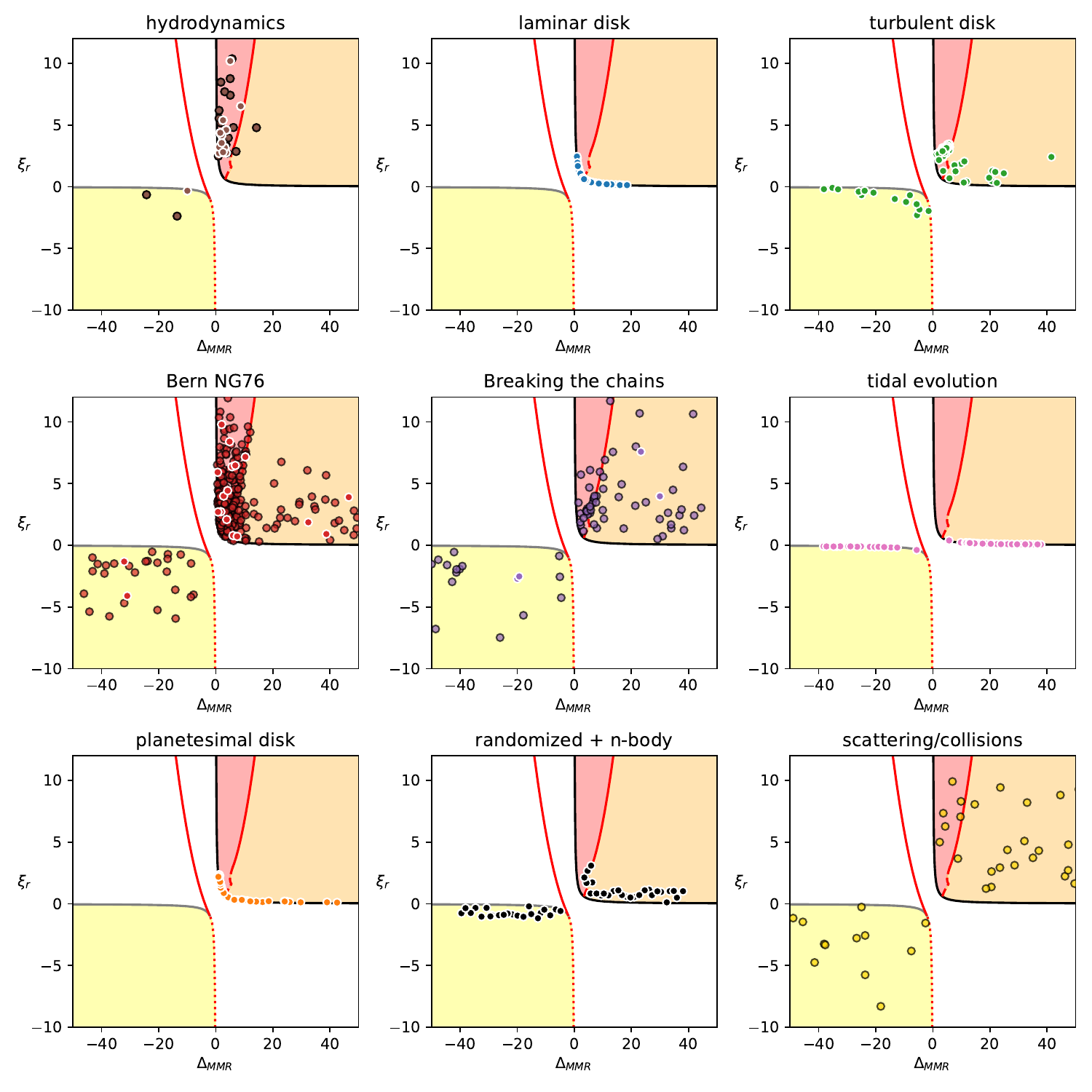}
\caption{Representation in the SFMR of all the pairs near first-order MMR in the final state of given runs. 
The hydrodynamics panel shows simulations from \cite{Ataiee2021}. The laminar disk and turbulent disk simulations are from \cite{GoBa2023}. Bern NG76 shows outcome of the Bern model run NG76 \citep{Emsenhuber2021,Burn2021}, while the `breaking-the-chains' panel shows outcome of three different runs of \cite{Izidoro2022,Shibata2025}. The tidal evolution and planetesimal disk simulations also come from \cite{GoBa2023}. The `randomized+ n-body' panel are system initialized with random period ratios and evolved to account for resonant repulsion. The`scattering + collision' panel correspond to simulation by \cite{Ishida&Kokubo2026}. Black circles indicate that the pair has neighbouring planets, while white circles indicate that the pair is isolated.}
\label{fig:SFM_compsim} 
\end{center}
\end{figure*}

 In this section, we explore how simulations that account for different physical processes yield different distributions in the ($\Delta_{\rm MMR}$, $\er$) plane. We note that most of the simulations in the database are not meant to quantitatively reproduce the entire observed population: although some of the simulations, such as the Bern model or the breaking-the-chains runs aim at reproducing the diversity of at least part of the exoplanet populations, others explore only a small range of planetary parameters. Nonetheless, these simulations give us an idea of what part of the resonant parameter space can be favoured by a given physical process. Future additions to the databases will enable more robust conclusions.

\textbf{Hydrodynamical simulations:} We start with the 2D hydrodynamical simulations from \cite{Ataiee2021}, an example of which is shown in Fig.~\ref{fig:hydro}. In that 2-planet simulation, the inner planet is trapped at the disk inner-edge, while the outer planet migrates inward, resulting in a decrease of the period ratio. The pair crosses the 2:1 MMR without being captured, while it is captured in the subsequent 3:2, 4:3 and 5:4 MMRs. The right panel shows the evolution through the 4:3 MMR: the pair arrives from $P_2/P_1>4/3$, with $\er \sim 0$, and follows the primary fixed point. Getting closer to the exact commensurability, the resonant eccentricity increases, up until the eccentricity damping balances out the effect of convergent migration. Then the pair undergoes over-stable libration \citep{GoSc2014,DeCoLa2015}, which leads to an increase of $\Delta \er$. Once the pair crosses the separatrix, $\er$ is damped and the pair migrates toward the next MMR \citep{DeBa2015}. The onset, timescale, and possible limit-cycle of  over-stability depend on the disk surface density and viscosity \citep{Afkanpour2024}. 
In this example, the simulation finishes as the pair is captured in the 5:4 MMR, with a slow increase of $\Delta \er$. The final positions of the simulated pairs and triplets of all the hydrodynamics runs are shown in the top-left panel of Fig.~\ref{fig:SFM_compsim}. The resulting pairs are predominantly in the resonant area, but not necessarily near the fixed points.\footnote{We note however that the disk is still present at the end of the simulation. The pair could therefore be closer to the fixed points if the remaining disk torque was taken into account \citep{BatPet2023}.}

\textbf{Laminar disk:} We continue with the 1D laminar disk simulations of \citet{GoBa2023} shown in the top-middle panel of Fig.~\ref{fig:SFM_compsim}. This run simulates the evolution of a pair of planets wide of the 3:2 MMR, with the effect of the disk modeled by a migration and an eccentricity damping timescale. These simulations reproduce what we typically expect from smooth convergent migration 
in absence of over-stable libration: a capture in MMR near the primary fixed point, with a final position that depends on the relative strength of the dissipative terms, deeper into the resonance for stronger migration or less eccentricity damping.
Based on this, the pair can end up in the resonant domain or not. However, we note that for pairs near the fixed points, there is no notable change in their trajectories when the phase space becomes resonant ($\delta=0$).

\textbf{Turbulent disk:} The turbulent simulations of \citet{GoBa2023} (top-right) add a stochastic force of varying amplitude to the laminar model. This results in pairs that are more spread-out with respect to the resonance, typically with non-zero distance from the fixed points, as well as resonant pairs that are roughly uniformly distributed in libration amplitude $\Delta \er$.

\textbf{Population synthesis:} The next two panels are the population synthesis Bern NG76 and the Breaking-the-chains runs described in \S\ref{sec:databases}. They use a disk model similar to the laminar disk simulations, but the disks evolve over time and the simulations span tens to a hundred million of years, the n-body simulations continuing after most of the disk is cleared.
In these simulations, pairs of planets are rarely isolated, and often form compact subsystems of 3+ planets that can still be represented pairwise in the SFMR \citepalias[see][]{GRSW1}.
 In these simulations, some of the pairs remain relatively close to the primary fixed points, albeit with a typically larger amplitude of libration $\Delta \er$; this can be due to the aforementioned overstable libration, or to a slow destabilization of a chain, as we will discuss below. 
Here again, pairs trapped by the resonance can either be in the resonance-trapped area, or in the upper area, as there is no sharp transition far from the separatrix \citep[e.g.][]{DeBa2015,Wang2026}.   
In these simulations, there are also numerous pairs in the lower and upper areas with relatively large absolute values of $\er \lesssim 5$. 
These are the remnants of broken resonant pairs through post-disk planet-planet scattering \citep{PuWu2015,Mills2016,Izidoro2017}: when a chain breaks after most of the proto-planetary disk dispersal, the planets exit the resonance, and can undergo further scattering/collisions but retain a significant resonant eccentricity. In absence of eccentricity damping, these systems remain non-negligible eccentricities. 
In both the breaking-the-chains runs and the Bern model, post-disk instabilities occur during the n-body integration, but the fraction of resonant pairs in the Bern model at the end of the simulation is much larger than the observed one ($25\%$ for the run NG76 after $100$ Myr). However, this fraction is still significantly evolving toward the end of the $100$ Myr of evolution and may approach the observed value at Gyr ages, see \S\ref{sec:time}. 

\textbf{Tidal evolution:} The next set of simulations shows the effect of tides, on the bottom right panel of Fig.~\ref{fig:SFM_compsim}. Planetary tides damp the eccentricity of planets, hence $|\er|$. Pairs initially inside the resonance follow the primary fixed points toward smaller $\er$, hence larger $\Delta_{\rm MMR}$. In some instances, the amplitude of libration $\Delta \er$ can increase, possibly leading to a crossing of the separatrix and a system ending near the secondary fixed points \citep{Delisle2014}. If the separatrix is not crossed, the pair typically ends up near the primary fixed points for $\Delta_{\rm MMR}>0$. If the planets are embedded in a resonant chain of first order resonances, even when only the innermost planets are subject to tides, it has the effect of moving each subsequent pair of planets along the primary fixed points, although in that case the dynamics can be complex \cite[e.g.][]{Papaloizou2015,Papaloizou2018,Charalambous2023}. The final state of pairs affected by planetesimal disks, not shown here, is similar to the tidally evolved systems. 

\textbf{Planetesimal disk:} The final state of the planetesimal disk run is in between the laminar disk and tidal run, with all pairs near the equilibria but without the occasional pair going interior to the resonance. The authors noted that retaining pairs near the resonance required some fine-tuning and that longer-term integrations tend to push pairs far from commensurability \citep{GoBa2023,Ghosh2023}.

\textbf{N-body interactions:} The last two runs shown in Fig. \ref{fig:SFM_compsim} are dominated by N-body interactions. In the `randomized + n-body' panel, isolated pairs were initialized symmetrically around the resonance \citep{GoBa2023}, with 0.01 eccentricities and randomized angles. This is a subset of the tidal evolution run, for which the damping timescale was longer than the integration duration, used here to show the effect of resonance repulsion. The `scattering + collisions' panel shows the result of simulations initialized with 15 protoplanets initially outside of resonances and with similar initial eccentricities \citep[see \S\ref{sec:databases},][]{Kokubo+2025}. These simulations typically finish with systems of 3 or 4 planets. Both the `randomized + n-body' and `scattering + collisions' runs end up with rather similar distribution in the upper and lower area, and few resonance-trapped pairs that are spread across the resonant domain. The final rescaled resonant eccentricities are, however, much larger for the pairs that were initially embedded in higher multiplicity systems.       

\begin{figure}
\begin{center}
\includegraphics[width=0.99\linewidth]{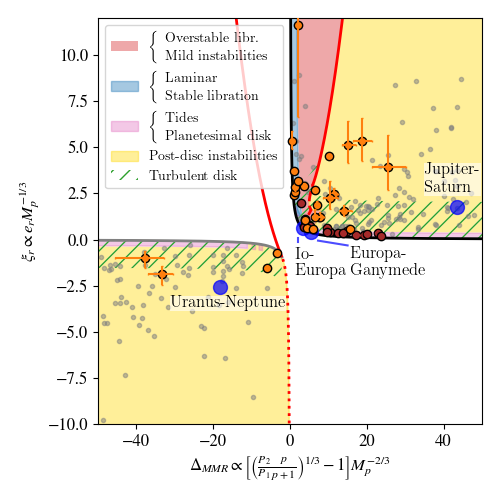}
\caption{
Representation in the SFMR of all the pairs near first-order MMR in the database (see \S\ref{sec:selection}). Brown markers show pairs robust to the mass-eccentricity degeneracy with an inner planet at less than 5 days, or part of a resonant chain whose innermost planet is at less than 5 days. Orange markers show the rest of the robustly characterised pairs. Grey dots show pairs that are either non robust, or not checked for robustness, see \citetalias{GRSW1}. The shaded areas indicate what physical process typically results in pairs observed in a given part of the phase space, see \S\ref{sec:simu}.
}
\label{fig:SFM_obstheory} 
\end{center}
\end{figure}

We summarize the information on the typical area of the $(\Delta_{\rm MMR},\er)$ plane that is reached by a given physical process in Fig.~\ref{fig:SFM_obstheory}.  %

\subsection{Comparison of observed and simulated populations}
\label{sec:scomparison}

In Fig. \ref{fig:SFM_obstheory}, we show the projection of all selected observed pairs on the SFM. The orange and brown marker shows the pairs that were found to be robust to the mass-eccentricity degeneracy \citep[][]{Boue2012,Lithwick2012}. The scarcity of robust characterization far from the resonance-trapped domain in the lower or upper areas is consistent with the mass-eccentricity degeneracy discussed in \citetalias{GRSW1}. Pairs inside and near the resonance-trapped domain are expected to not be affected by that degeneracy. However, the absence of pairs close to the separatrix is expected for stability reasons. 

One of the strongest accumulations of pairs is near the primary fixed points at low $\er$ in the upper area. Most of these pairs have either an inner planet whose orbital period is less than 5 days, or are part of a resonant chain whose innermost planet has an orbital period of less than 5 days \citepalias{GRSW1}, and the position of these pairs in the SFMR matches that of the `tidal evolution' panel of Fig. \ref{fig:SFM_compsim}. In order to enable the comparison with simulations in which tides are not modeled, we discard the pairs, and resonant chains, with an inner period at less than 5 days for the rest of the study.

The simulations do not necessarily reproduce the relative fractions of planet pairs in the different zones of SFMR phase space. Moreover, some configurations generate higher TTVs and some are more likely to be robustly measured. We cannot easily account for this observational bias in our heterogeneous database. So, to quantitatively compare the simulated systems to observed ones, we consider three zones of the phase space---the resonant domain (\S\ref{sec:reso}), the upper area (\S\ref{sec:xlibphase}), and the non-resonant upper and lower area (\S\ref{sec:fracextint})---and compare the simulated and observed SFMR parameters within those regions.
We provide the Kolmogorov-Smirnov (KS) test metrics and p-values between each simulation set and the observations in Table~\ref{tab:summary} for the four regions.


\subsubsection{Distribution in the resonant domain}
\label{sec:reso}

\begin{figure}
\begin{center}
\includegraphics[width=0.99\linewidth]{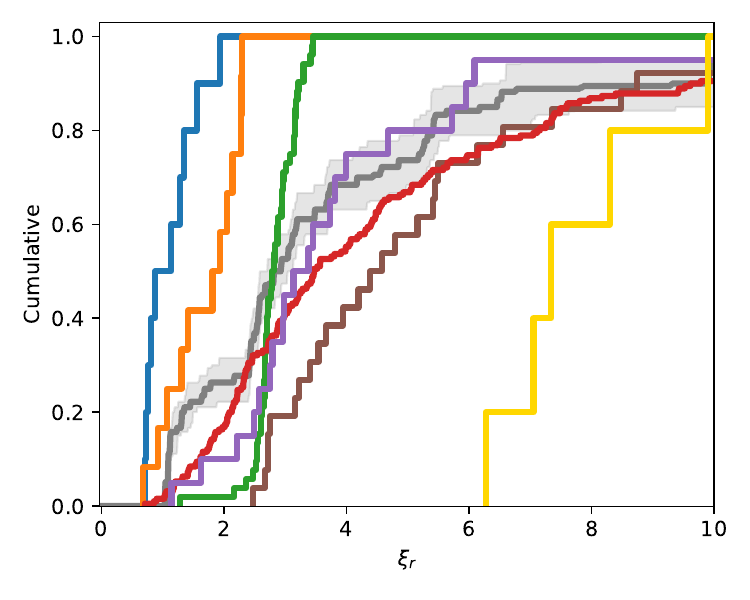}\\
\includegraphics[width=0.99\linewidth]{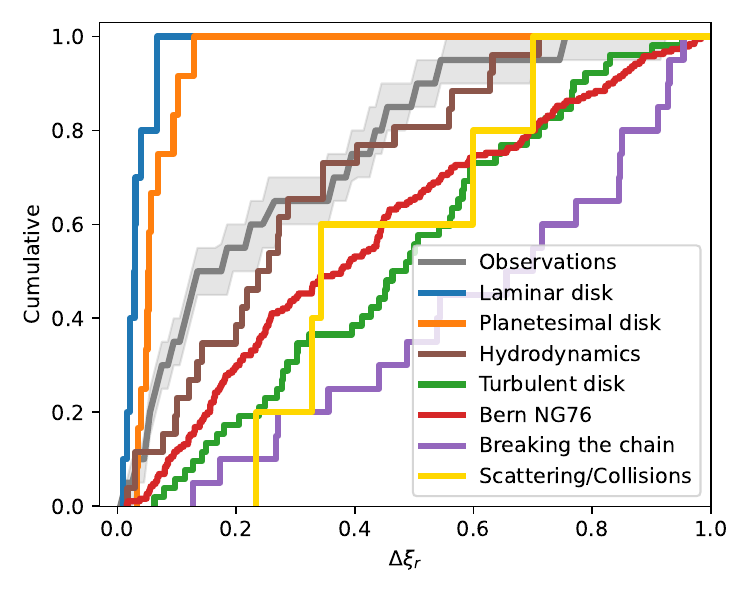}
\caption{
Distributions in the resonance-trapped domain (cyan in Fig. \ref{fig:SFM_obstheory}). The observed population is based on 300 samples for each of the 17 characterized pairs with $P_{\rm in}>5$ days, including those that were not tested for robustness. Top: distribution of $\er$, bottom: distribution of $\Delta \er$. 
}
\label{fig:xlib_reso} 
\end{center}
\end{figure}

We first study pairs in the resonance-trapped domain, which is the red area in Fig.~\ref{fig:SFM_anarea}. 
The mass-eccentricity degeneracy does not affect this area of the phase space \citepalias{GRSW1}; we therefore consider all characterised pairs.
Fig.~\ref{fig:xlib_reso} shows the distributions of $\er$ and of $\Delta\er$ for observed and simulated systems in this region. 
The 17 observed pairs are roughly uniformly distributed over $\er$ in the [1-5] range, while low $\Delta \er$ are favoured, with a median at $\sim 0.2$.

The physical processes predict radically different distributions of both quantities in the resonant domain: laminar and planetesimal disks predict the whole population to be within $\er \lesssim 2$ and $\Delta \er<0.1$, which is quite different from the observed population. The turbulent simulation has a roughly uniform distribution in $\Delta \er$, and $\er\sim 3$ for the whole run. 
The hydrodynamic simulations have a distribution of $\Delta \er$ in the resonant domain that is quite close to the observed resonant population; their distribution of $\er$ is qualitatively similar to the observations but consistently overestimates it by $1-2$.
The Bern model NG76, in red, follows roughly the correct distribution for both quantities. These distributions do not evolve significantly between 10 and 100 Myr of post-disk evolution. The breaking-the-chains runs roughly predict the correct distribution of $\er$, but $\Delta \er$ is too large compared to the observed population.
The scattering/collisions run generates pairs that are very different from the observed population, with large $\er$ and large departure from the primary equilibrium.

\subsubsection{Distribution of $\Delta \er$ in the upper area}

\label{sec:xlibphase}
\begin{figure}
\begin{center}
\includegraphics[width=0.99\linewidth]{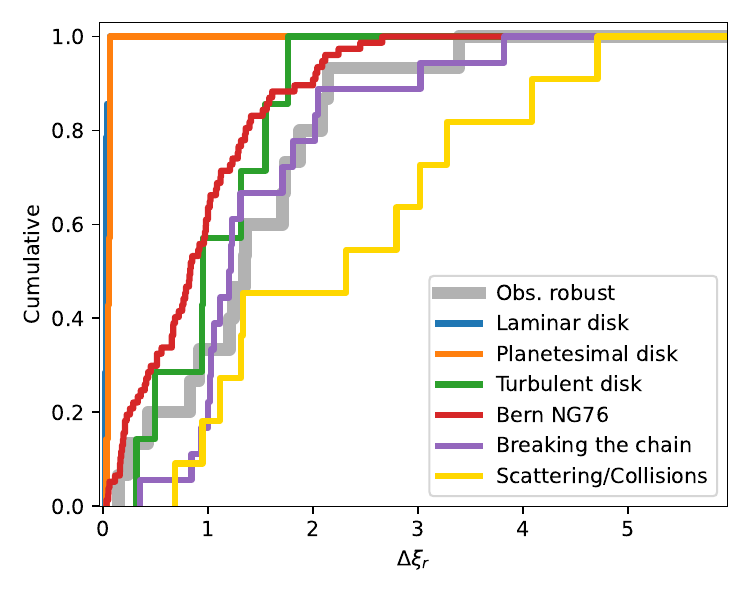}
\caption{Cumulative distribution of the distance to the primary fixed points $\Delta \er$ in the upper area. The observed population is based on all robust pairs with $P_{\rm in}>5$.
}
\label{fig:xlib_phases} 
\end{center}
\end{figure}

The position of planet pairs in the upper and lower areas differs significantly across our tested simulations.
In this section, we focus on the upper area as an example because it contains the well-known excess of planet pairs with period ratio above the exact commensurability. Figure \ref{fig:xlib_phases} shows that laminar disks, as well as dissipation through remaining planetesimal disks, predict all pairs to end up very close to the fixed points, with $\Delta \er \sim 0$. The simulation of turbulent disks predicts a somewhat uniform distribution of $\Delta \er$ in the 0--1.5 range, with the overall distribution, and upper limit, depending on the strength of the stochastic force added to the model. The population synthesis models predict a dearth close to the fixed points at $\Delta \er \sim 0$, a peak at $1 - 1.5$, and a heavy tail toward larger $\Delta \er$; while the scattering/collision run results in a somewhat uniform distribution in the $[1\!-\!5]$ range. 
We do not compare to the hydrodynamic simulations here or in the next section because they produce very few non-resonant pairs.

Regarding observations, most of the upper area is dominated by the mass-eccentricity degeneracy, although there are well constrained pairs close to the resonant domain ($\smash{\Delta_{\rm MMR}=10.39_{-3.84}^{+7.73}}$). In \citetalias{GRSW1}, we showed that pairs with larger $\Delta_{\rm MMR}$ ($\smash{=14.8_{-7.9}^{+14.1}}$) could have a $\Delta \er$ distribution consistent (albeit with a somewhat heavier tail toward large values) with the robustly-characterized subsample using their TTV phase distribution \citep{Lithwick2012}. 


The robustly-characterized observed population has a somewhat uniform distribution of $\Delta \er$ in the [0--2] range, in good agreement with the turbulent disk run, as well as with both population synthesis models, although the distribution of the Bern model is slightly shifted toward lower values.

\subsubsection{Distribution of $\Delta_{p+1:p}$ for non-resonant pairs}
\label{sec:fracextint}

\cite{Fabrycky2014} showed that the upper part ($P_2/P_1 >(p+1)/p$) of first order MMRs was more densely populated than the lower part ($P_2/P_1 <(p+1)/p$) in the Kepler sample. Here, we revisit this question with the additional knowledge of whether a pair is actually trapped in resonance.

\begin{figure}[h!]
\begin{center}
\includegraphics[width=0.95\linewidth]{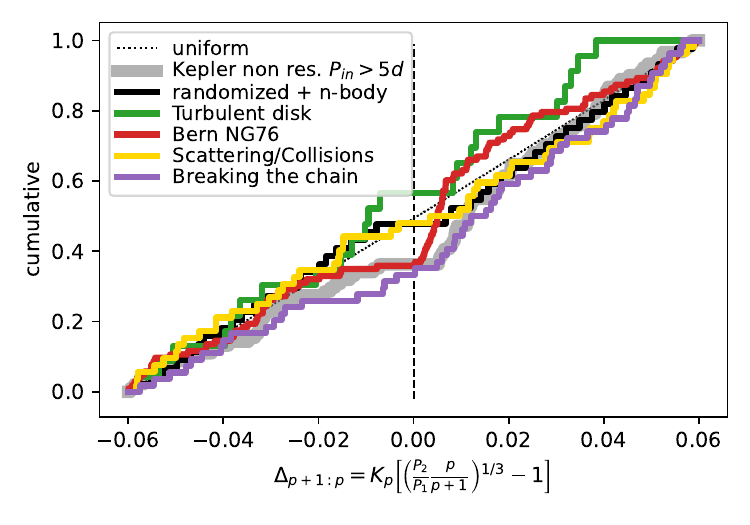}\\ \vspace{-.35cm}
\includegraphics[width=0.95\linewidth]{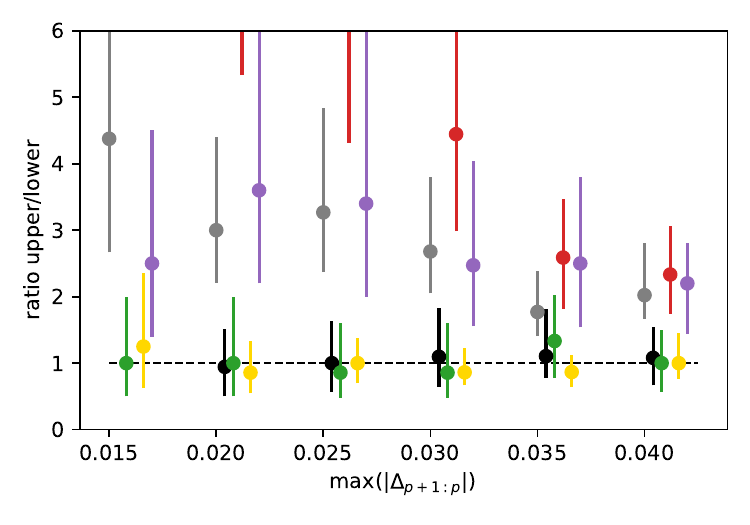}\\ \vspace{-.35cm}
\caption{Distribution of pairs above and below the exact commensurability for observed and simulated systems. 
The constant $K_p$ depends only on the chosen MMR (eq. \ref{eq:Dpp1p}). 
%
(a) comparison to models for pairs in the vicinity of the 3:2 and 2:1 MMRs. Pairs in the resonance-trapped domain were removed (cyan area in Fig. \ref{fig:SFM_obstheory}). The cumulatives are normalized by the total number of pair in each case. (b) ratio of the number of pairs above and below the commensurability for a given maximum value of $|\Delta_{p+1:p}|$. 
}
\label{fig:fracextint} 
\end{center}
\end{figure}

To have the largest possible sample, we consider all Kepler confirmed multis verifying $P_{\rm in}>5$ d, $P_{2}<70$ d and $1.4<P_2/P_1<2.5$. This period ratio range is chosen to select only the 3:2 and 2:1 MMRs, for which the range of $\Delta_{p+1:p}$ \citepalias[which is symmetric,][]{GRSW1} is wide enough to encompass part of the continuum, i.e., where the slope of the cumulative is the same as a uniform period ratio distribution. We cross check them with the database and discard pairs that are in the resonance-trapped domain. $\Delta_{p+1:p}$ is preferred over $\Delta_{\rm MMR}$ in this section since we do not have robust masses for most of the Kepler pairs. The cumulative distribution of this quantity centered around 0 is shown in grey in Fig \ref{fig:fracextint} (a), and is made of 104 pairs. It shows the characteristic dearth of pairs below commensurability, and over-abundance above commensurability, mentioned in the Introduction.

To compare with the observed population, we first consider the turbulence run \cite{GoBa2023} (green); the randomized N-body run of \cite{GoBa2023} (black), and the scattering/collision run of \cite{Ishida&Kokubo2026} (yellow).
All these result in distributions consistent with the uniform period ratio distribution shown by the dotted black line. 

The considered Bern NG76 and `breaking-the-chains' population synthesis runs, in contrast, end up with a pronounced asymmetry between the lower and upper partition similar to the observed population, with fewer pairs for small negative values of $\Delta_{p+1:p}$, and an increase once the commensurability is passed. The rate of this increase near 0 is sharper for the Bern model NG76, than in the observed population. However, 
when exploring the temporal evolution of the Bern NG76 run, it appears that this distribution is still evolving towards the end of the 100 Myr simulations. 
Longer integrations are required to obtain the final distribution for that population, a point that we will return to in \S\ref{sec:time}. In contrast, the `breaking-the-chains' runs, made of chains that break early, have a distribution very close to the observed one. Panel (b) of Fig. \ref{fig:fracextint} quantifies the ratio of pair above and below the commensurability within a given distance. The turbulent disk and randomized runs are consistent with a uniform distribution, while the breaking-the-chains and Bern NG76 runs are consistent with the observations for $\max (|\Delta_{p+1:p}|) \geq 0.025$, despite the latter having too many near-resonant pairs (Fig. \ref{fig:Bern_fraction}) and pairs that are too close from commensurability at 100 Myr.
This asymmetry between the upper and lower areas is therefore consistent with the aftermath of broken resonances.

\begin{table*}[!h]
\centering
\caption{Summary of how each physical process compares with four of the trends presented in \S\ref{sec:comparison}. In each cell, the Kolmogorov–Smirnov test is reported: the first value is the maximum distance between the cumulatives, the second is the p-value, and the number in parenthesis is the number of pairs. {\color{cadmiumgreen}green} lower than 0.2, {\color{orange}orange} lower than 0.4 and {\color{red}red} larger than 0.4. $^*$ This distribution is still evolving after $100$~Myr.
See the text for more details. }\label{tab:summary}
\begin{tabular}{c|c c c c}
     & $\er$ & $\Delta \er$   & $\Delta \er$ & $\Delta_{p+1:p}$ \\
    & reso. & reso. & upper &  \\
    \hline
\# of obs.&(17)&(17)&(15)&(104)\\
\hline
Laminar&\textcolor{red}{0.82|$1.0\times10^{-4}$}(10)&\textcolor{red}{0.71|$1.6\times10^{-3}$}(10)&\textcolor{red}{1.0|$2.6\times10^{-8}$}(14)&-(0)\\
Planetesimals&\textcolor{red}{0.82|$2.9\times10^{-5}$}(12)&\textcolor{red}{0.50|0.037}(12)&\textcolor{red}{1.0|$1.2\times10^{-5}$}(7)&-(0)\\
Hydrodynamics&\textcolor{orange}{0.40|0.060}(26)&\textcolor{orange}{0.26|0.41}(26)&-(0)&-(0)\\
Turbulent disk&\textcolor{red}{0.41|0.018}(52)&\textcolor{red}{0.51|$1.3\times10^{-3}$}(52)&\textcolor{orange}{0.27|0.78}(7)&\textcolor{orange}{0.22|0.34}(20)\\
Bern NG76&\textcolor{cadmiumgreen}{0.16|0.78}(190)&\textcolor{orange}{0.38|0.015}(190)&\textcolor{orange}{0.37|0.046}(78)&\textcolor{orange}{0.30|$4.3\times10^{-4}$}(81)$^{*}$\\
breaking-the-chains&\textcolor{cadmiumgreen}{0.15|0.94}(20)&\textcolor{red}{0.65|$3.0\times10^{-4}$}(20)&\textcolor{orange}{0.22|0.73}(18)&\textcolor{cadmiumgreen}{0.098|0.89}(45)\\
Scattering/Collisions&\textcolor{red}{0.82|$4.3\times10^{-3}$}(5)&\textcolor{red}{0.59|0.094}(5)&\textcolor{red}{0.48|0.076}(11)&\textcolor{orange}{0.23|0.086}(40)\\
\end{tabular}
\end{table*}

\section{Discussion}

\label{sec:discussion}
\subsection{Implications for planet formation}
Now that we have tested published simulations against data, our goal is to gain additional insight into the physics of the planet formation process. To this end, we discuss the relative importance of a number of physical effects that were included in at least one simulation we tested.

\textbf{Planetary tides:} We found in \citetalias{GRSW1} a strong correlation between the resonant eccentricity $|\er|$ and the orbital period of the inner planet of a resonant configuration. In particular, for robustly or precisely characterised pairs, a strong dichotomy appears at $P_{\rm in} \sim 5$ d, as was also noted by \cite{HaLi2017}. When comparing $P_{\rm in} < 5$ d systems to the simulated systems, we interpret this correlation as evidence for the tidal evolution of this sub-population of planets \citep{Delisle2014,HaLi2017}. 
 
We then discarded observed systems with $P_{\rm in} < 5$ d for the rest of the study to ensure that we were not comparing tidally evolved systems to runs that do not simulate tides.

\textbf{Disk migration:} Orbital migration in the protoplanetary disk remains the most effective mechanism to produce planets in and near resonance \citep{LePe2002,CreNe2008}. While pure dissipative evolution of two planets without resonant overstability or post-disk scattering is disfavoured at a population level, hydrodynamic simulations of migration in low-viscosity disks closely match the distribution in the resonant domain, especially in $\Delta\er$. The slight discrepancy in $\er$ is likely due to the disk aspect ratio used in the \citet{Ataiee2021} simulations, which sets the equilibrium value \citep{Pichierri2018}. The fact that population synthesis runs, which compute migration and damping timescales from standard Type I prescriptions and a background disk model, are also a satisfactory match in the resonant domain supports the validity of this approach and implies that the ratio of migration to damping obtained from the disk model \citep{TaTaWa2002,TaWa2004} is reasonably accurate. However, both models overestimate the excitation away from resonant equilibrium and lack a small population of chains that remain near the equilibria. This could be due to a missing dissipative effect or an excess of planets in the resonant chains. An alternative possibility is that being very close to the equilibrium could be required for stability over Gyr.

We do not find evidence in favor of the importance of stochastic kicks from turbulent eddies \citep{Rein2012}. This process does not consistently produce the period ratio asymmetry near resonance and is a poor match to the observed diversity in $\er$, although the latter may be due to a single planet mass used in the highly simplified simulations of \citet{GoBa2023}. Furthermore, the breaking-the-chains and Bern model simulations neglect turbulence yet produce overexcited resonant systems. An Occam's razor argument would suggest that turbulence is therefore not necessary. A more thorough analysis of resonant chains could provide a quantitative upper limit on turbulence strength for specific systems, as was performed by \citet{Huhn2021} for Kepler-223.

\textbf{Planetesimal disk:} Scattering of small particles \citep{ChatterjeeFord15,Ghosh2023} carries planet pairs along the resonant equilibrium curve in a manner similar to tidal dissipation. However, unlike tides, this effect does not explain the strong trend of $|\er|$ with the orbital period of the inner planet. In fact, debris belts are likely more prevalent at wider orbital separations where collisional grinding is less destructive. 
We note that there is a continuum in perturber mass between planetesimal scattering and ``ping-pong'' scattering of Moon--Mars-sized embryos, the latter of which appears to be more able to leave near-resonant pairs with nonzero $\Delta\er$ \citep{Wu2024}.

\textbf{Post-disk planet-planet interactions:}
The runs accounting only for orbital migration produced almost no pairs outside of resonance, since pairs that crossed resonances without capture or exited resonances due to overstable libration migrated and got captured in more compact MMRs.\footnote{Note that our tested simulations did not include certain models of planet-disk interactions that can produce non-resonant systems, such as inviscid disks \citep{McNally2019} and movement of the disk cavity \citep{Liu2017,Pichierri2024}.} This implies that subsequent physical processes need to take place to explain the mostly non-resonant observed population. Out of the models available in the database, the Bern model NG76 runs \citep{Emsenhuber2021} and the breaking-the-chains runs \citep{Izidoro2022,Shibata2025} display trends broadly coherent with the observed population in most of the points of focus of our analysis for both resonant and non-resonant pairs.

The dominant driver of post-disk evolution in these models is gravitational interactions between planets: resonant chain breaking, secular evolution, planet-planet scattering. It is important to note that scattering alone does not explain the population we consider. While scattering can occasionally produce planet pairs trapped in resonance, it does not explain the frequent occurrence of several resonances in the same system \citep{Dai2024}. Moreover, $\er$ following scattering is essentially uniformly distributed within stability limits, much larger than the observed values \citep{Yee2021}; and the observed clustering of resonant systems near the fixed point $\Delta\er=0$, strongly suggesting formation in dissipative conditions, is not reproduced. Finally, our scattering/collision run, in which the planets were initialized outside of resonance, does not reproduce the over density of pair above commensurability. So, while dynamical instabilities and scattering seem common in compact systems, the systems remaining near MMR show evidence that they did not experience a destabilization that completely erased the resonant initial conditions.

\subsection{Did most multiplanet systems start in resonance?}
\label{sec:time}
\begin{figure}
\begin{center}
\includegraphics[width=0.99\linewidth]{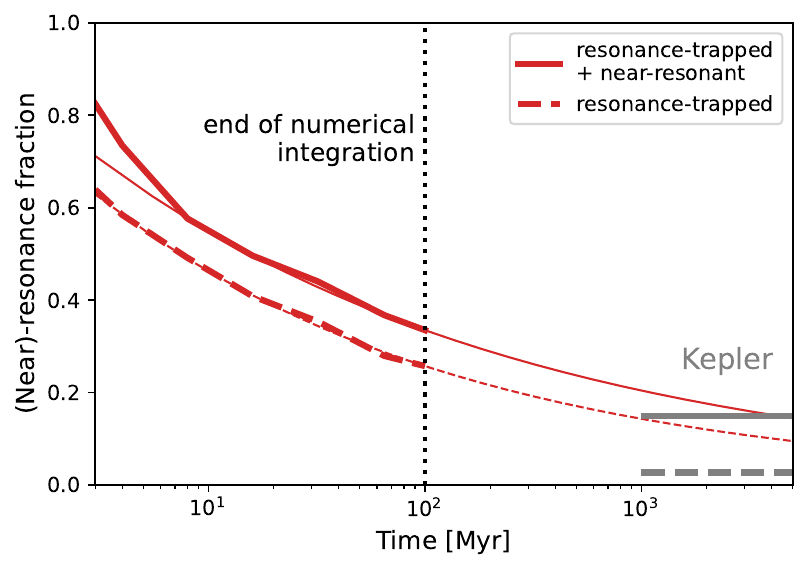}
\caption{Fraction of resonance-trapped (red in Fig.~\ref{fig:SFM_obstheory}) and resonance-trapped + near-resonant (satisfying $|\Delta_{p+1:p}|<0.025$) in the Bern model run NG76. The thick lines show the evolution during the $100$ Myr post disk evolution available in the \texttt{resonantstate} databases, while the thin lines show our projections to 5Gyr. The grey lines show the fraction for the confirmed Kepler multis with $P_{\rm in}>5$ days. Some of the Kepler resonant pairs might be counted as near-resonant if they are not in the database.
}
\label{fig:Bern_fraction} 
\end{center}
\end{figure}

We now return to the two main theoretical modeling problems raised in the introduction: the rarity of resonant systems and the large offsets from commensurability for near-resonant ones. Models that exhibit a breaking-the-chains-type evolution have been found to reproduce the near-resonant fraction, at least for certain sets of simulation parameters \citep{Izidoro2021}. Nevertheless, a major caveat is that it is only computationally feasible to numerically integrate several hundred Myr of dynamical evolution given the short orbital timescales, but most of the transiting planet sample orbits Gyr-old stars \citep{Berger2020}. \cite{Izidoro2021} demonstrated that continuing these integrations longer can lead to a much higher fraction of unstable systems more consistent with the Kepler sample. \citet{GoPe2026} extrapolated the instability rate for a similar set of simulations and showed that it could be consistent with the observed decay as a function of time found by \cite{Dai2024}.
Repeating this process for the Bern model runs, we found that both the resonant and the (near-)resonant populations evolve roughly linearly in log-fraction vs log-time from a few Myr after the disk dispersal to the simulation end at $100$ Myr, 
following 
\begin{equation}
\log_{10}f_{\rm res} = -0.25 \log_{10}(t/\textrm{yr}) + 1.45
\end{equation}
for the resonant fraction and 
\begin{equation}
\log_{10}f_{\rm (near-)res} = -0.21 \log_{10}(t/\textrm{yr}) + 1.24
\end{equation}
for the combination of resonant and near-resonant, respectively (Fig.~\ref{fig:Bern_fraction}).
Extrapolating these laws to $5$ Gyr yields a fraction of $\sim 17\%$ of combined resonant and near-resonant systems ($\Delta_{p+1:p}<0.025$) and $9.9\%$ of resonant systems. The former is about the same as the observed pairs with $\Delta_{p+1:p}<0.025$ in Kepler, while the fraction of resonant pairs is $\sim 3$ times larger than the observed one. 
Although our metric of distance to the resonance is not exactly the same as \cite{Dai2024}, we also recover a fraction of near-resonant systems over time broadly consistent with their observational results. However, the fraction of young systems that are resonance-trapped probably remains more abundant in simulations than in observations. It is not clear whether this is due to a limitation of the model or TTV-related biases against detecting resonant systems.

Upon exiting the resonance, planets are not uniformly distributed in phase space. Figure~\ref{fig:fracextint} shows that non-resonant planet pairs in population synthesis models preferentially lie wide of resonance, with about 60\% of pairs having $\Delta_{p+1:p} > 0$. This is not a guaranteed consequence of planet-planet scattering or the definition of $\Delta_{p+1:p}$ because the scattering-collision runs are symmetric around 0. Since in the latter runs planets do not originate in resonance, this appears to be a distinct phenomenon associated with chain breaking.

\begin{figure}
\begin{center}
\includegraphics[width=0.99\linewidth]{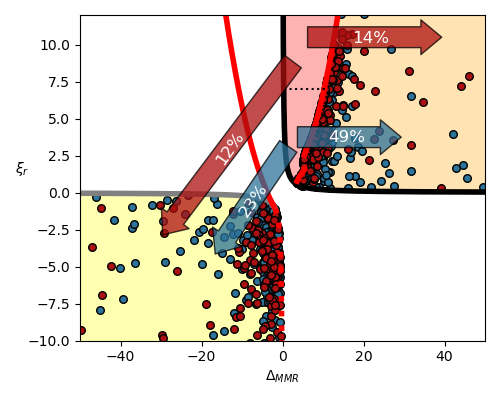}\\
\caption{Position of pairs of neighbouring planets in the SFMR the moment the planets exit resonance (the Bern model NG76), between 4 and 100 Myr. The color red (resp. blue) shows the pairs that had a rescaled resonant eccentricity above (resp. below) 7 before exiting the resonance. The percentage in the arrow indicate the fraction of pairs that exited following a given path. Only 20\% of these pairs end up near an MMR at 100 Myr. Out of these, 88\% of the pairs that originally exited to the upper area are still found in the upper area of an MMR at 100 Myr, while only 37\% of those that exited toward the lower area are still found in the upper area of an MMR at 100 Myr.   
}
\label{fig:Bernbreak} 
\end{center}
\end{figure}

Analyzing this temporal evolution further, it appears that the fraction of broken pairs that exit the resonance towards the upper or lower areas depends on how deeply pairs are typically embedded in the resonance: pairs that exit the resonance from large $\er$ have a roughly equal chance to exit above or below commensurability, while pairs at low eccentricity favour the upper area, possibly due to the asymmetry of the phase space and the absence of a separatrix on that side for $\er<3$; see top panel of Fig.~\ref{fig:Bernbreak}. Although most of the pairs strongly evolve after they first exit resonance, the pairs that remain close to the resonance tend to remain on the side they broke. This is of particular interest since the value of $\er$ that is reached by a pair of planets during convergent migration depends on the ratio between the eccentricity damping and migration timescales, hence the properties of the proto-planetary disk.

The fraction of near, but not resonance-trapped pairs remains relatively constant over time in the Bern model integrations (Fig.~\ref{fig:Bern_fraction}). This suggests that there is a roughly steady-state population of near-resonant systems that comprises the abundance of systems with $\Delta_{p+1:p}$ just above 0 seen in Fig.~\ref{fig:fracextint}. This population is gradually lost to violent instabilities but replenished from resonant systems slowly diffusing out of resonance. We show in Appendix~\ref{ap:time} that its distribution of $\Delta\er$ is approximately stationary. The breaking-the-chains runs, which are more dynamically evolved than the Bern model, have a weaker peak at that location. Indeed, the Bern model chains that break before 50 Myr have a significantly flatter distribution around $\Delta_{p+1:p}=0$ that resembles the breaking-the-chains runs and the observations (Fig.~\ref{fig:fracextint_time}). We caution that the exact shape of these distributions depends on the somewhat arbitrarily chosen boundaries of the resonance-trapped domain.

As a whole, these results are consistent with the hypothesis that both the near-resonant excess of planet pairs wide of commensurability and the typical non-resonant systems are the consequence of disruptions of resonant chains. 
We also find evidence that some systems can undergo a two-stage disruption, first moving slightly away from resonance and then sustaining a violent instability \citep{Choksi2026,HaWu2026}. Mild breaking can reproduce the asymmetry near resonance, and unlike dissipative processes such as tides, explains the moderate eccentricities of near-resonant planets. We do not exclude tides from operating in the most close-in systems (recall that Figs.~\ref{fig:xlib_phases} and \ref{fig:fracextint} exclude pairs with $P_{\rm in}<5$ d, which \citetalias{GRSW1} shows are likely tidally evolved), but not relying on tides in general also evades some puzzling results, including requiring unexpectedly strong tidal dissipation \citep{Lee2013} and considerable resonant offsets in some of the youngest systems \citep{Dai2024}. Upcoming measurements of architectures of near-resonant systems as a function of stellar age from PLATO will offer stronger constraints on the breaking-the-chains hypothesis and refine the relevant timescales.

\subsection{Alternative scale-free variables and comparisons}
\label{sec:scalefree}
In this paper, we projected near-resonant planet pairs into the SFMR in order to combine simulations and observations that span a wide range of periods, masses, and resonant indices. While first-order resonant dynamics are indeed determined uniquely by $\Delta_{\rm MMR}$ and $\er$, the physics that creates, sculpts, and breaks resonances may scale differently. For example, under the standard assumption that migration and eccentricity damping rates are proportional to planet mass $m$, the equilibrium eccentricity is nearly independent of $m$ and period $P$ \citep{Pichierri2018}. In contrast, eccentricities following planet-planet scattering scale as $m^{1/3}P^{1/3}$ \citep{Kokubo+2025}. Since $\er \propto e m^{-1/3} P^0$, these processes predict a weakly mass and period-dependent distribution of $\er$. 
Likewise, resonant overstability depends on the resonant index $p$ and thus the final resonance reached by a migrating pair is related to the disk conditions \citep{Kajtazi2023,Batygin2026}. 
This is not likely to be a severe problem because our chosen simulations are mostly intended to reproduce typical planet masses and periods, and the database is dominated by relatively uniform hot super-Earth/sub-Neptune systems, but future work can slice the population further. 
Our particular choice of parameters was primarily motivated by the constraints provided by TTV modeling, but we note that this was a major topic of discussion at the 2025 Geneva Resonant State Workshop and other choices are possible, especially in the context of 3+ planet systems and more unusual architectures.

\begin{figure*}[h]
\begin{center}
\includegraphics[width=0.99\linewidth]{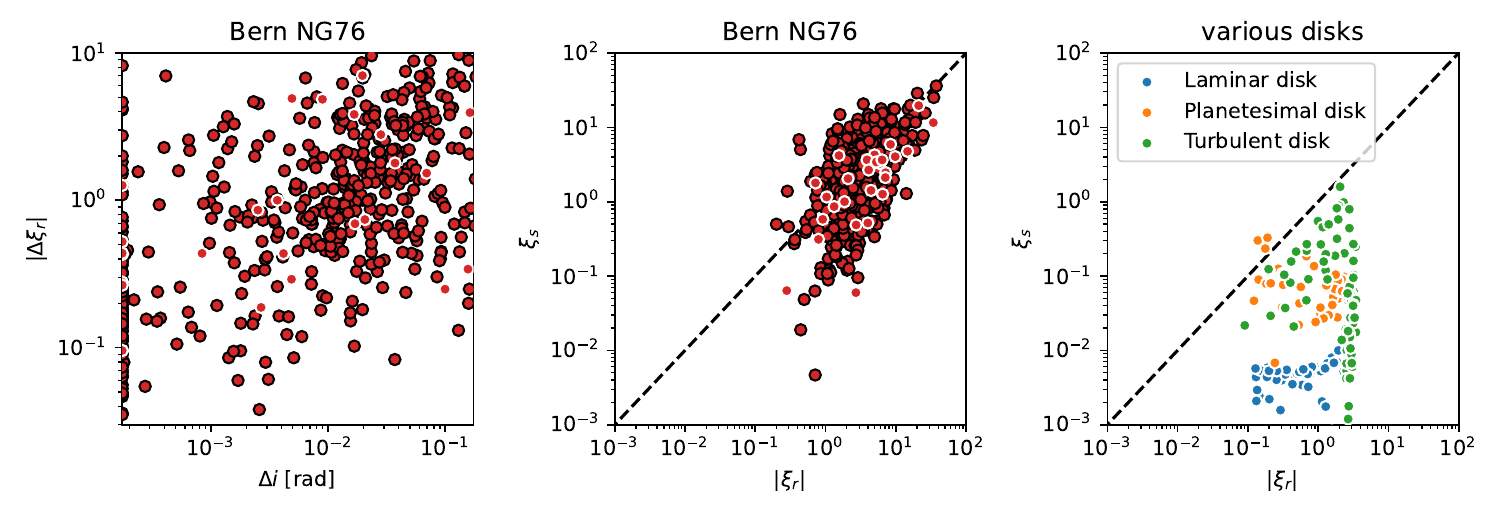}
\caption{Correlation between various scale-free variables. The circle indicates if a pair is isolated (white) or part of a larger subsystem (black). For the right panel, different colors indicate different disks from \cite{GoBa2023}.  }
\label{fig:mutinces} 
\end{center}
\end{figure*}

Other quantities of interest include the rescaled secular eccentricities $\es$ and the mutual inclinations $\Delta i$, which are complementary observables to the ones extensively discussed in this paper. 
In Fig. \ref{fig:mutinces}, we show how these quantities correlate with the quantities of the SFMR in a few simulations. In the Bern model, $\Delta i$ and $\es$ correlate positively with $|\Delta \er|$ and $|\er|$, which can be interpreted as a sign of dynamical heating in a system. The correlation between $|\er|$ and $\es$ also strongly depends on the simulated physical processes, with near-zero $\es$ expected for resonant capture under idealized migration and eccentricity damping \citep[the so-called apsidal corotation resonances, see][]{HaPa2020}. Comparison to observations is harder for these quantities: the inclinations were not treated homogeneously for the different analyses in the database (assumption of coplanarity, fixed ascending nodes, etc.), and the rescaled secular eccentricity is rarely constrained \citepalias{GRSW1}.
Even so, the minimum value of the mutual inclination can be estimated for any pair even without TTV solutions \citep[e.g.,][]{Fabrycky2014} through an estimation of the planets' impact parameters. Small mutual inclination is not affected by planetary tides for close-in systems and therefore could be a useful marker of past gravitational instabilities. In addition, there is a sub-population of single planets with significant TTVs \citep[a hundred or more in Kepler: ][]{Holczer2016,Yahalomi2024} that could be used to estimate the distribution of mutually-inclined near-resonant pairs \citep[e.g.,][]{Millholland2021}. However, single planet systems with TTVs are often difficult to interpret because the inversion problem is intrinsically degenerate \citep{NeMo2008}, with similar TTV signals arising from perturbers near multiple distinct resonances \citep[e.g.,][]{YaKi2026,LaWi2026}. We leave these comparisons to observations for future work.

\section{Conclusion}
\label{sec:conclusion}

In this work, we used the \texttt{resonantstate} database, which collects community contributions of observed multiplanet system architectures as well as thousands of simulations of planet formation, and applied an analysis framework based on the second fundamental model of resonance (SFMR). We compared the dynamical state of observed systems -the rescaled resonant eccentricity and distance to resonance \citepalias{GRSW1}- to simulations in order to identify the physical processes responsible for producing trends in orbital architectures. We present the following conclusions:
\begin{itemize}
    \item As found in \citetalias{GRSW1}, close-in systems ($P_\textrm{in}\!<\!5$ d) lie very close to the resonant equilibrium \citep{HaLi2017} and somewhat offset from exact commensurability \citep{Delisle2014letter}, consistent with the effect of tidal dissipation within the planets and in the same regime as the Galilean satellites Laplace chain. We discard systems with $P_\textrm{in}\!<\!5$ d in the remaining analysis because most simulations did not include the effects of tides.
    \item For planet pairs trapped in resonance, simulations of migration and resonance capture in the disk typically reproduce the observed rescaled resonant eccentricities, both in hydrodynamical codes and n-body integrations using prescriptions of migration and eccentricity damping, corroborating these models. The reproduction of the proximity to the fixed point depends on the run, and may depend on the (in)stability mechanisms taking place.
    \item Near-resonant pairs have a distribution in period ratios and rescaled resonant eccentricities that is inconsistent with dissipative evolution but consistent with post-disk instabilities, although a phase of capture in resonance seems required to generate the excess of systems wide of commensurability compared to narrow of it. We interpret this to be due to the typical resonant eccentricity reached during the planets' resonant capture and the asymmetry of the resonant phase space.  
\end{itemize}

This study, combined with Paper I, corroborates the breaking-the-chains hypothesis for a significant fraction of the observed population. Simulations that underwent gentle breaking of resonances best reproduce the near-resonant observed pairs with an inner planet at more than 5 days of orbital period. 
For closer-in systems, a similar evolution followed by the effects of planetary tides, 
provides a fitting initial condition for these pair to reach their observed departure from commensurability (\citetalias{GRSW1}). 
We recall that, although our papers focused on near-resonant pairs, the breaking-the-chains process is able to reproduce the period ratio distribution of the whole Kepler sample \citep{Izidoro2017,Izidoro2021}. Indeed a similar pathway from a resonant chain of giant planets has been invoked to explain the architecture of the outer Solar System \citep{Morby-et-al-2007,Thommes2008}.

 However, we do not exclude the possibility that other physical processes could reproduce on their own the discussed observables, or provide important contributions for a better quantitative match between simulations and observations. In particular, synthetic populations that do not undergo a phase of resonant capture would be very valuable.
 We invite additional contributions\footnote{Please contact adrien.leleu@unige.ch.} of simulations of formation and evolution of multi-planetary systems to be added to the database, whether they aim to replicate a large segment of the exoplanet population or an individual system, to enable future comparisons.

The PLATO mission \citep{PLATO} is expected to discover hundreds of multi-planetary systems showing TTVs and, using a combination of TTV and ground-based RV data, provide constraints on their orbital architectures. This new sample will have better-understood observational biases and detection limits compared to our existing highly heterogeneous database. Furthermore, PLATO should measure ages to a precision of $<10\%$ of exoplanet host stars, which will enable future searches for age-related trends in the quantities we have studied in this work. Like Kepler and TESS, the PLATO sample will be dominated by Gyr-old stars and hence most observed trends with age will fall outside of most numerical models, including all used in this work, that focus on the first few hundred Myr or less.

\bibliographystyle{aa}
\bibliography{biblio}

\appendix

\nolinenumbers

\begin{acknowledgements}
AL, YE, JC, JK and MR acknowledge support of the Swiss National Science Foundation under grant number TMSGI2\_211697. This work has been carried out within the framework of the NCCR PlanetS supported by the Swiss National Science Foundation under grants 51NF40\_182901 and 51NF40\_205606. SH acknowledges support by the Natural Sciences and Engineering Research Council of Canada (NSERC), funding references CITA 490888-16 and RGPIN-2020-03885. MG acknowledges support from the ERC project No. 101019380 "HolyEarth.". National Aeronautics and Space Administration Exoplanet Research Program (NASA XRP) grant 80NSSC21K1111. This work was performed in part by the Virtual Planetary Laboratory Team, which is a member of the NASA Nexus for Exoplanet System Science, and funded via the NASA Astrobiology Program ICAR Grant 80NSSC23K1398. DR is supported by the National Science Foundation under Award No. 2143195. JK acknowledges support from the Spanish Research Agency of the Ministry of Science, Innovation and Universities (AEI-MICIU) under grant `Contribution of the IAC to the PLATO Space Mission' with reference PID2023-149439NB-C41. KM was supported by JSPS KAKENHI grant No 25K07387. GP acknowledges support from the European Union (ERC Starting Grant DiscEvol, project number 101039651) and from Fondazione Cariplo, grant No. 2022-1217. Views and opinions expressed are, however, those of the author(s) only and do not necessarily reflect those of the European Union or the European Research Council. Neither the European Union nor the granting authority can be held responsible for them. LBo acknowledges support from CHEOPS ASI-INAF agreement n. 2019-29-HH.0. ACMC acknowledges support from FCT, Portugal, through the CFisUC project UID/04564/2025 (with DOI identifier 10.54499/UID/04564/2025). FD acknowledges support by NSF CAREER Award 2441950. IGJ acknowledges support of National Science and Technology Council in Taiwan under grant NSTC 114-2112-M-007-029. ACP has been supported by the French government, through the $ \mathrm{UCA^{J.E.D.I.}} $ Investments in the Future project managed by the National Research Agency (ANR) with the reference number ANR-15-IDEX-01 MW acknowledges support of the National Natural Science Foundation of China (grant No.124B2058). TGW acknowledges support from the University of Warwick and UKSA. DAY is supported by a Juan Carlos Torres Postdoctoral Fellowship at the Massachusetts Institute of Technology. This work was granted access to the HPC resources of MesoPSL financed by the Region Île-de-France and the project Equip@Meso (reference ANR-10-EQPX-29-01) of the  programme Investissements d’Avenir supervised by the Agence Nationale pour la Recherche. LCY acknowledges support of National Science and Technology Council in Taiwan under grant NSTC 114-2115-M-007-007.
\end{acknowledgements}

\section{Temporal evolution of Bern model NG76}
\label{ap:time}

\begin{figure}
\begin{center}
\includegraphics[width=0.99\linewidth]{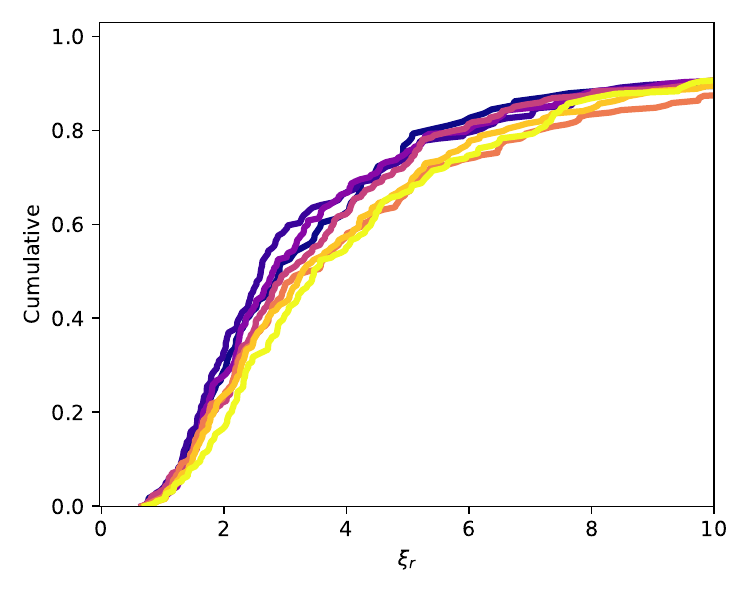}
\includegraphics[width=0.99\linewidth]{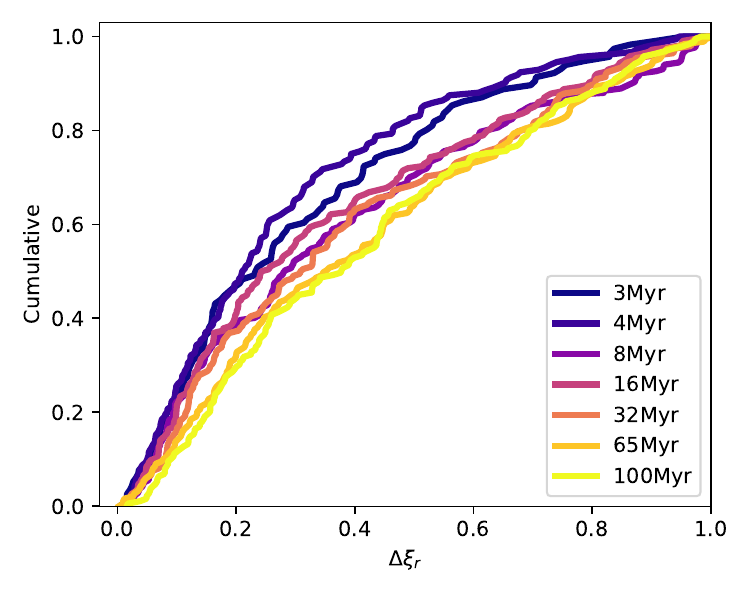}
\caption{Excitation of the rescaled resonant eccentricities in the resonance-trapped area for the NG76 run of the Bern model at several epochs, see Fig.~\ref{fig:xlib_reso}. 
}
\label{fig:xlib_reso_time} 
\end{center}
\end{figure}

\begin{figure}
\begin{center}
\includegraphics[width=0.99\linewidth]{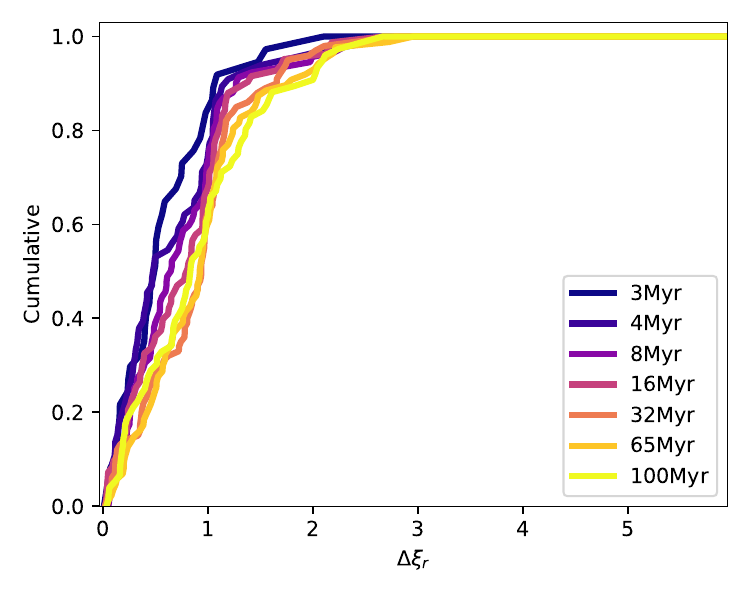}
\caption{Excitation of the rescaled resonant eccentricities in the upper area for the NG76 run of the Bern model at several epochs, see Fig.~\ref{fig:xlib_phases}. 
}
\label{fig:xlib_ext_time} 
\end{center}
\end{figure}

\begin{figure}
\begin{center}
\includegraphics[width=0.99\linewidth]{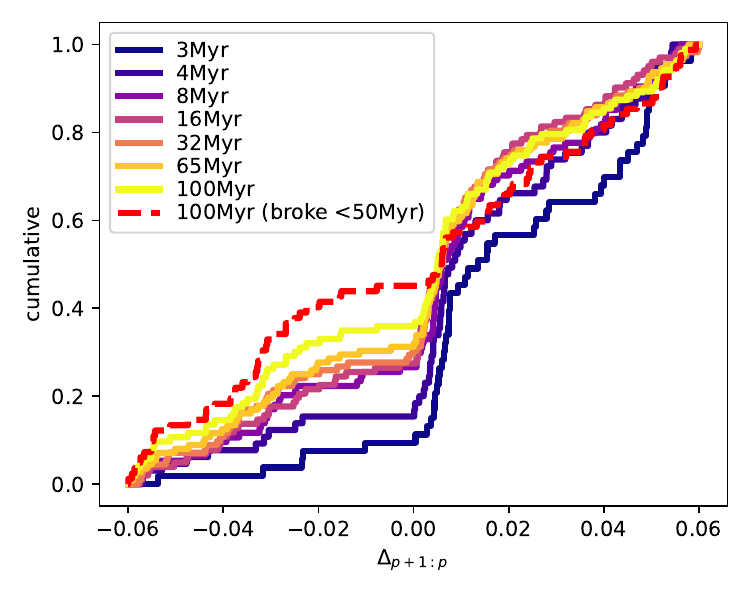}
\caption{Distribution of pairs on either side of commensurability for the NG76 run of the Bern model at several epochs, see Fig.~\ref{fig:fracextint}.
}
\label{fig:fracextint_time} 
\end{center}
\end{figure}

In this appendix, we show how the various trend discussed in the paper evolve over the 100 Myr of the Bern NG76 run. The distribution of the resonance-trapped pair, discussed in \S \ref{sec:reso}, is shown in Fig. \ref{fig:xlib_reso_time}; the distribution of $\Delta \er$ of pairs above commensurability from \S \ref{sec:xlibphase} is shown in Fig. \ref{fig:xlib_ext_time}, and the distribution in $\Delta_{p+1:p}$, discussed in \S \ref{sec:fracextint}, is shown in Fig. \ref{fig:fracextint_time}. The trends discussed in \S \ref{sec:reso} and \S \ref{sec:xlibphase} seem to be reaching a steady state toward 100 Myr, while the distribution in $\Delta_{p+1:p}$ is still evolving. In particular, the red dashed line in Fig. \ref{fig:fracextint_time} shows that the subset of pairs that broke before 50 Myr have a distribution in $\Delta_{p+1:p}$ significantly different at 100 Myr than the full population.     

\end{document}